\documentstyle[preprint,revtex]{aps}

\newcommand{\bqq}{\begin{equation}}
\newcommand{\eqq}{\end{equation}}

\begin{document}
\draft

\begin{title}
Ultraviolet and soft X--ray photon--photon elastic scattering in an electron
gas.
\end{title}

\author{R. Marto\v{n}\'{a}k, E. Tosatti\cite{et}}

\begin{instit}
International School for Advanced Studies (S.I.S.S.A.) \\
Via Beirut 2--4, 34014 Trieste, Italy
\end{instit}


\begin{abstract}
We have considered the processes which lead to elastic scattering
between two far ultraviolet or X--ray photons while they propagate inside a
solid, modeled as a simple electron gas. The new ingredient, with
respect to the standard theory of photon--photon  scattering in
vacuum, is the presence of low--energy, nonrelativistic
electron--hole excitations. Owing to the existence of two--photon
vertices, the scattering processes in the metal are predominantly of
second order, as opposed to fourth order for the vacuum case. The main
processes in second order are dominated by exchange of
virtual plasmons between the two photons. For two photons of similar
energy $\hbar \Omega$, this gives rise to a cross section rising like
$\Omega^2$ up to maximum of around $10^{-32}$~cm$^2$, and then decreasing
like $\Omega^{-6}$. The maximal cross section is found for  the photon
wavevector $k \sim k_{F}$, the Fermi surface size, which typically
means a photon energy $\hbar \Omega$ in the keV range. Possible experiments
aimed at checking the existence of these rare but seemingly measurable
elastic photon--photon scattering  processes are discussed, using in
particular intense synchrotron sources.
\end{abstract}

\section {\bf Introduction}

Photon--photon scattering processes are well known in vacuum.
The lowest order perturbation theory diagrams are shown on
Fig.\ref{fig:1}. The corresponding scattering amplitude and cross
section,  first calculated by Karplus and Neumann \cite{karpneu}, were
later thoroughly  investigated by de Tollis \cite{detoll}, and are
extensively reviewed in Lifshitz and Pitaevski\v{i}'s book \cite{alp}.
The cross section of the process in vacuum turns out to be very small,
of the order of magnitude of $10^{-30}cm^{2}$ at the threshold for the
production of real $e^{-}-e^{+}$ pairs in the intermediate state,
which  corresponds to an incident photon energy of $\hbar\Omega = m
c^{2} \sim  0.5$~MeV. For energies $\hbar \Omega \ll m c^{2}$, the
cross section falls as $\sigma \sim ( \frac{\hbar \Omega}{m
c^{2}})^6$. It is thus extremely small  in the eV range, where
powerful laser beams are available, or even in the  keV range, where
strong synchrotron sources are now feasible.
The physical reason is clear: if the energy of photons lies too far below
the "energy gap" $2 m c^{2}$, the virtual intermediate $e^{-}-e^{+}$
pairs are hard to excite, and this greatly cuts the probability of the
process. In the end, this makes the process very difficult to pursue
experimentally -- albeit not impossible \cite{zavat1}, \cite{zavat2}.

One is thus led to the obvious idea that, if instead of vacuum, the
photons were embedded in a material medium, where pair excitations
could be created with a smaller or zero energy gap, then the
probability of photon-photon scattering  in such a system might be
enhanced.

Condensed matter physics provides us with countless examples of
systems with  this property: electron--hole pairs replace
electron--positron pairs, and this reduces excitation energies from
MeV to the eV or meV range. For example, in a semiconductor the
intermediate state involves pairs excited  across the
semiconducting band gap.

Third order nonlinear optical
susceptibilities, which in principle involve the same diagrams of
Fig.\ref{fig:1}, have for example been well characterized, both
experimentally \cite{garito}, and  theoretically \cite{chang},
\cite{weikang}, for the quasi--one--dimensional  prototypical
semiconductor polyacetylene.

The focus of this paper will be however on the elastic scattering
cross section among very high energy photons, which makes the
problem radically different from those studied so far.

To make a start, we would like to study photon--photon scattering inside the
simplest and most typical condensed system, for example that of a
simple metal. Of  course, even simple metals  have in real life a
number of complications. The periodic potential brings about a
multi--band problem. Due to the multiplicity of bands, interband
transitions may occur, etc. However we shall restrict in this paper
to the contribution of {\it intraband} transitions,
right at the Fermi surface, and these processes can be modelled
with a single parabolic band. Moreover, in the keV photon energy range (which
will turn out to be the most interesting case) virtually all
systems, including insulators, can often be treated as free  electron metals
to a very good approximation.

A second complication is that in a metal, or more generally in a solid
or  liquid, photons are heavily absorbed. Their  energy decays into
electron--hole pairs, and finally into heat. Of course, this  fact may
represent a practical obstacle at getting strong photon fields inside
an actual material. Yet, for example, one can think of using a very
intense  evanescent photon field at a surface in a reflection
experiment, or use intense, far ultraviolet and X--ray beams from
a synchrotron radiation source, tuned so that the the absorption cofficient is
small enough to be neglected.

A third relevant issue might be  the difficulty of separating elastic
from inelastic photon--photon processes, when the process takes place
inside a metal. Since there is no gap, a shower of soft electron--hole
pairs is likely to accompany the photon--photon collision, and it
could be difficult in practice to separate this inelastic part
from the {\it elastic} process we are after.

Postponing further discussion of some of these, and other
related issues to a later section of this paper, we will now proceed
with the actual calculation of the elastic photon--photon cross
section in an electron gas, so as to come up with predictions for
orders of  magnitude and angular dependences. This calculation
is not simply a generalization of the classic
calculation of \cite{alp} from  vacuum to a uniform electron gas.
The nonrelativistic
nature of the electron gas allows for new processes, which do not
exist in vacuum. There are new energy scales, the Fermi
energy $E_{F}$ and the plasma energy $\hbar \omega_{p}$, which are
absent in vacuum.

Anticipating our later results, we will end up with numerical
values of the cross section, which are again exceedingly small - we
estimate for a potassium target a peak value of $10^{-32}$~cm$^2$,
somewhat smaller than that in vacuum.
However, {\it the peak is now in  the keV photon frequency range}.
Since ultraviolet and soft X--ray photon sources are so much more
intense than $\gamma$--ray sources, the possibility that our
calculations might not remain academic seems real. Moreover, our
estimated  photon--photon cross section in the ultraviolet is
$\sim 10^{-35}$~cm$^2$, which might be measurable using medium power
lasers. At this stage, it is not clear to us whether such measurements
are going to be feasible, or worth doing, or whether the present calculation
will remain an academic exercise. We simply would like to point out that
the new photon-photon processes exist, and that they are in principle finite
and measurable.

As announced, we shall take the uniform, single--band non--relativistic
electron gas (EG)
as our  model for the metal. The model is completely specified by the
electron mass $m$ and the Fermi energy $E_{F}$. In section 2 we
identify the simplest possible  photon--photon scattering processes,
which now appear in the second, third and fourth order of the
perturbation theory (as opposed to just fourth order in vacuum). In
later sections, we choose two of these processes and calculate them
out in detail.  First, in section 3 we deal with the 2nd order
processes, which are the largest, and consist essentially of exchange
of plasmons between the photons. These  processes do not exist in
vacuum. In section 4 we then investigate the 4th order process,
formally the same square diagram  of Fig.\ref{fig:1}.
Now, however, positrons are replaced by
holes in  the Fermi sea, which modifies both the form and the
magnitude of the resulting  scattering amplitude. The resulting
fourth--order cross section, not  surprisingly, turns out in the end
to be many orders of magnitude smaller than  that of the 2nd order
processes, even at very high frequencies (where they could in
principle have come closer). We therefore identify plasmon exchange as
the dominant mechanism for the photon-photon scattering in the EG. In
section 5  we summarize our results, and discuss briefly the possibilities of
experimentally  observing the proposed photon--photon elastic
scattering processes.

\section {\bf Possible scattering processes}

Before moving on to the actual calculation, we wish to stress some
points, which make the situation of photons inside the EG different from that
in  vacuum. First of all, a photon in an electron gas is a
well--propagating mode  only at frequencies much larger than the
plasma frequency $\omega_{p}$.  Approaching that frequency from above,
photons become dressed, absorption increases, until below $\omega_{p}$
photons do not propagate anymore, and get  absorbed within a
wavelength or so. In an actual experimental situation, a beam  from an
external source penetrates into the metal, and the scattered,
reflected or transmitted radiation, which leaves the metal, is
detected. In these  conditions, what actually scatters inside the
metal, are the {\it dressed} photons, as they enter through the surface.
To predict the outcome of such experiment rigorously, one would have to
perform the calculation including photon dressing, plus the effect of the
surfaces. Both represent nontrivial complications, which we shall not
endeavor to consider at this stage. Therefore, we simply assume here free
photons propagating in an infinite bulk electron gas. We shall restrict our
treatment  to frequencies $\Omega$ higher than  $\omega_{p}$
\bqq
\hbar \Omega > \sim \hbar \omega_{P} \, ,
\eqq
and completely neglect the effects of photon dressing, and surface effects.
This will allow us to use essentially
the same formalism as in the case of  vacuum.

Since we want to treat the electrons as non--relativistic, we shall
also  restrict the energy $\hbar \Omega$ of our photons from above
\bqq
\hbar \Omega \ll m c^{2} \, . \label{eq:nonrel1}
\eqq
We shall also assume, trivially, that the Fermi
energy of the electron gas (imagining typical metal density) satisfies
\bqq
E_{F} \ll m c^{2} \, , \label{eq:nonrel2}
\eqq
obviously satisfied in practice.

The starting point is the non-relativistic hamiltonian for electrons
interacting with electromagnetic field in the Coulomb gauge
\begin{eqnarray}
H &=& \int d^{3}x \Psi ^{\dagger}(x) {{1} \over
{2m}}\,(\vec{p} + {{e} \over {c}}  \vec{A}(x))^{2}\, \Psi (x)
+ \sum_{\vec{k} \lambda} \hbar\Omega_{\vec{k}}
a^{\dagger}_{\vec{k} \lambda}  a_{\vec{k} \lambda}
\nonumber\\
&+&  {{1} \over {2}}\, \int \int d^{3} x_{1} d^{3} x_{2}
\Psi ^{\dagger}(x_{1}) \Psi ^{\dagger}(x_{2}) U(x_{1}-x_{2}) \Psi
(x_{2}) \Psi (x_{1}) \, ,
\end{eqnarray}
the electron charge being $-e$. Here
$\Psi^{\dagger}(x)$ creates an electron at point $x$ (spin indices are
omitted for simplicity) and  $a^{\dagger}_{\vec{k} \lambda}$ creates a
photon of wavevector $\vec{k}$ and polarization $\lambda$. The
$\vec{A}$ operator is
\bqq
\vec{A}(\vec{x}) = \sum_{\vec{k} \lambda}
\left({{4 \pi \hbar c^{2}} \over  {2 \Omega_{\vec{k}} V}}\right)^{1/2}
e^{i\vec{k} \vec{x}} \vec{e}_{\vec{k} \lambda} (a_{\vec{k} \lambda} +
a_{-\vec{k} \lambda}^{\dagger}) \, ,
\eqq
where $V$ is the
normalization volume.  The total hamiltonian can be split into the
following terms
\bqq
H = H_{0} + H_{Coul} + H_{rad} + H_{i1} + H_{i2} \, ,
\eqq
where
\bqq  H_{0} =  \int d^{3}x \Psi ^{\dagger}(x) {{1}
\over {2m}} \vec{p}^{2} \Psi (x)
\eqq
and
\bqq  H_{rad} =
\sum_{\vec{k} \lambda} \hbar \Omega_{\vec{k}} a^{\dagger}_{\vec{k}
\lambda} a_{\vec{k} \lambda}
\eqq
are the free fields,
\bqq
H_{Coul}
= {{1} \over {2}} \int \int d^{3} x_{1} d^{3} x_{2}  \Psi
^{\dagger}(x_{1}) \Psi ^{\dagger}(x_{2}) U(x_{1}-x_{2}) \Psi (x_{2})
\Psi  (x_{1})
\eqq
is the Coulomb interaction between the electrons and
\bqq
H_{i1} = {{e} \over {m c}} \int d^{3} x \Psi ^{\dagger}(x)
\vec{A}(x). \,  \vec{p} \Psi (x) \label{eq:hi1}
\eqq
and
\bqq
H_{i2} = {{{e}^{2}} \over {2 m c^{2}}} \int d^{3} x
\Psi ^{\dagger}(x) \Psi(x) \vec{A}(x) . \vec{A}(x) \label{eq:hi2}
\eqq
are the interaction
terms linear and quadratic in $\vec{A} (x)$, respectively.

Now to have a non-zero S-matrix element $S_{fi}$ between the initial
photon state containing 2 photons $\vec{k}_{1},\vec{k}_{2}$ and the
final state with two photons $\vec{k}_{1}^{'},\vec{k}_{2}^{'}$, we must
have in the S--operator $S(-\infty,+\infty)$ a term with at least
four $\vec{A}$ operators
\bqq
S_{fi} = \langle \Phi_{FS} \mid \, \langle \Phi_{ph} \mid \,
a_{\vec{k}_{1}^{'}} \,  a_{\vec{k}_{2}^{'}} S(-\infty,+\infty)
a_{\vec{k}_{1}}^{\dagger} \,  a_{\vec{k}_{2}}^{\dagger}\, \mid
\Phi_{ph} \rangle \, \mid \Phi_{FS} \rangle \, ,
\eqq
$\mid \Phi_{ph} \rangle$ and $\mid \Phi_{FS} \rangle$ being the
photon vacuum and the ground state of the
Fermi sea of electrons, respectively. The photon polarization index
$\lambda$ has been absorbed into $\vec{k}$. Elasticity of the process
is implied by the Fermi sea being left in its ground state at the end
of the  process.

For photon--photon scattering in vacuum, described by the relativistic
hamiltonian, where the interaction term is linear in $\vec{A}$, the
lowest order of perturbation theory giving nonzero contribution is the
4th \cite{karpneu}, \cite{detoll},\cite{alp}. The corresponding
processes are the three diagrams on  the Fig.\ref{fig:1}, differing
from each other by the assignment of the external legs to the vertices
of the square, plus three other diagrams which differ from  the former
ones just by the orientation of the internal fermionic loop.

In our case, however, there are more possibilities, because the interaction
term  $H_{i2}$ is quadratic in $\vec{A}$. The simplest
scattering  processes now appear in second, third and fourth order
of perturbation theory, and are characterized by the respective
scattering amplitudes
\begin{eqnarray}
& & S_{fi}^{(2)} = ({{-i} \over
{\hbar}})^{2} {{1} \over {2!}} \int_{-\infty}^{+\infty} \!\! dt_{1}
\nonumber \\
& & \int_{-\infty}^{+\infty} \!\! dt_{2} \,
 \langle \Phi_{FS} \!\mid\! \langle \Phi_{ph} \!\mid\!
a_{\vec{k}_{1}^{'}}  a_{\vec{k}_{2}^{'}} \,
T[H_{i2}(t_{1}) H_{i2}(t_{2})] \, a_{\vec{k}_{1}}^{\dagger}
a_{\vec{k}_{2}}^{\dagger} \!\mid\! \Phi_{ph} \rangle \!\mid\! \Phi_{FS}
\rangle \, , \label{eq:2order} \\
& & S_{fi}^{(3)} = ({{-i} \over {\hbar}})^{3} {{1} \over {3!}}
\int_{-\infty}^{+\infty} \!\! dt_{1} \, \ldots
\nonumber \\
& & \int_{-\infty}^{+\infty} \!\! dt_{3} \,
\langle \Phi_{FS} \!\mid\! \langle \Phi_{ph} \!\mid\!
a_{\vec{k}_{1}^{'}}  a_{\vec{k}_{2}^{'}} \,  T[H_{i1}(t_{1})
H_{i1}(t_{2})  H_{i2}(t_{3})] \, a_{\vec{k}_{1}}^{\dagger}
a_{\vec{k}_{2}}^{\dagger} \!\mid\! \Phi_{ph} \rangle \!\mid\! \Phi_{FS}
\rangle \, , \label{eq:3order} \\
& & S_{fi}^{(4)} = ({{-i} \over {\hbar}})^{4} {{1} \over {4!}}
\int_{-\infty}^{+\infty} \!\! dt_{1} \, \ldots \nonumber \\
& &\int_{-\infty}^{+\infty} \!\! dt_{4} \,
\langle \Phi_{FS} \!\mid\! \langle \Phi_{ph} \!\mid\!
a_{\vec{k}_{1}^{'}}  a_{\vec{k}_{2}^{'}} \,  T[H_{i1}(t_{1})
H_{i1}(t_{2}) H_{i1}(t_{3})  H_{i1}(t_{4})] \,
a_{\vec{k}_{1}}^{\dagger} a_{\vec{k}_{2}}^{\dagger} \!\mid\!
\Phi_{ph} \rangle \!\mid\! \Phi_{FS} \rangle \, , \label{eq:4order}
\end{eqnarray}
where all operators are now in the interaction picture.  In the
following sections we shall investigate two of the above processes,
$S_{fi}^{(2)}$ and $S_{fi}^{(4)}$ in actual detail, and evaluate their
scattering amplitudes.

Since we shall finally be interested in the cross sections of the
scattering  processes, we recall that the
usual scattering amplitudes  $M_{fi}$ satisfy the relations
\bqq
S_{fi} = \delta_{fi} + i (2 \pi)^4 \delta (P_{f}-P_{i}) T_{fi} \, ,
\label{eq:s}
\eqq
\bqq
T_{fi} = {{1} \over {\sqrt{2 \Omega_{\vec{k}_1}
V}}} \, \ldots \,  {{1} \over {\sqrt{2 \Omega_{\vec{k}_{2}^{'}} V}}}
M_{fi} \, , \label{eq:t}
\eqq
and that the differential scattering cross section
${{d\sigma} \over {do}}$ is related to $M_{fi}$ through the
equation
\bqq
{{d\sigma} \over {do}} = {{1} \over {64 \pi^{2} c^4}}  {
{\mid M_{fi} \mid^{2}} \over  {(\Omega_{1} + \Omega_{2} - c
(\vec{k}_{1}+\vec{k}_{2}).\vec{n})^2} } \, , \label{eq:dsdo}
\eqq
where $\Omega_{1,2}$ and $\vec{k}_{1,2}$ are frequencies and
wavevectors of  the incident photons and $\vec{n}$ is a unit vector in
the direction of the  solid--angle element $do$.

\section{\bf Second order scattering processes: plasmons}

In this section we shall deal with the processes which result
from the second order scattering amplitude (\ref{eq:2order}). In the end,
these will be the dominant contribution to the scattering, due to the
smallness of the fine structure constant $\alpha = {{e^2} \over {\hbar
c}} = {{1} \over {137}}$. Substituting for $H_{i2}$ from
(\ref{eq:hi2}) and using the fact that the $\Psi$ and $\vec{A}$
operators commute we get
\begin{eqnarray}
S_{fi}^{(2)} &=& ({{-i} \over {\hbar}})^{2} {{1} \over {2!}}
({{e^{2}} \over {2 m c^{2}}})^{2} \int d^{4} x_{1} \int d^{4} x_{2} \,
\langle \Phi_{FS} \!\mid
T[\Psi^{\dagger}(x_{1}) \Psi(x_{1}) \Psi^{\dagger}(x_{2})
\Psi(x_{2})] \mid\! \Phi_{FS} \rangle \nonumber \\
&\times& \langle \Phi_{ph} \!\mid
a_{\vec{k}_{1}^{'}} a_{\vec{k}_{2}^{'}} \, T[(\vec{A}(x_{1}) .
\vec{A}(x_{1})) (\vec{A}(x_{2}) . \vec{A}(x_{2}))] \,
a_{\vec{k}_{1}}^{\dagger} a_{\vec{k}_{2}}^{\dagger} \mid\!
\Phi_{ph} \rangle \, .
\end{eqnarray}
Using Wick's
theorem, we find that there are six diagrams, three of which are shown
on Fig.\ref{fig:3}. The three remaining diagrams differ from these
just by interchange of the vertices and thus yield exactly the same
contribution. Diagrams 1 and 2 of Fig.\ref{fig:3} can be regarded as
processes where the 4-momentum transfer between the photons is
accomplished via exchange of an electron-hole pair. Diagram 3
describes instead the creation of an electron-hole pair by two--photon
absorption and subsequent decay of the pair into two final photons.

Performing the usual algebra and introducing the electron Green's
functions $G^0$ we get for $M_{fi}^{(2)}$ from the bare diagrams on
Fig.\ref{fig:3} (including the factor of 2 from other 3 diagrams which
just cancels the ${{1} \over {2!}}$ factor)
\begin{eqnarray}
M_{fi}^{(2)} &=& ({{-i} \over {\hbar}})^{2} \, ({{e^{2}} \over {2 m
c^{2}}}) ^{2} \, (4 \pi \hbar c^{2})^{2}
\sum_{diagrams} (\vec{\epsilon}_{i1} . \vec{\epsilon}_{i2}) \,
(\vec{\epsilon}_{i3} . \vec{\epsilon}_{i4}) \, {{(-2 i)} \over {(2
\pi)^{4}}} \, \int d^{4}k \, G^{0}(k) G^{0}(k + p_{i}) \nonumber\\ &=&
-{{1} \over {\hbar}} \, ({{e^{2}} \over {2 m c^{2}}})^{2} \, (4 \pi
\hbar c^{2})^{2} \, \sum_{diagrams} (\vec{\epsilon}_{i1} .
\vec{\epsilon}_{i2}) \, (\vec{\epsilon}_{i3} . \vec{\epsilon}_{i4}) \,
\Pi^{0}(p_{i}) \, . \label{eq:bare}
\end{eqnarray}
In the last
expression we have introduced the usual Lindhard complex polarizability
function
$\Pi^{0}(p_{i})$ of the free electron gas \cite{fetter}, with $p_{i}$
for the i-th diagram determined by the 4-momentum conservation in the
vertex. The unit polarization vectors of the photons incident with the
vertex 1, resp. 2 of the i-th diagram have been denoted as
$\vec{\epsilon}_{i1}$, $\vec{\epsilon}_{i2}$ and
$\vec{\epsilon}_{i3}$, $\vec{\epsilon}_{i4}$, respectively.

To proceed further, we notice an important difference between
the interaction hamiltonians $H_{i1}$ and $H_{i2}$. In a homogeneous
medium, the electron-hole
pair created by a photon through $H_{i1}$ is transverse and therefore
cannot decay into a longitudinal Coulomb interaction. Hence, the
process of Fig.\ref{fig:4} has zero amplitude. If however, as in the
present case (Fig.\ref{fig:3}), the electron-hole pairs are created by
$H_{i2}$, this selection rule is absent, and there is nothing to prevent decay
of the pair and a
subsequent creation of another one due to Coulomb interaction.
Therefore we must renormalize the bare diagrams of Fig.\ref{fig:3}.
Within the random phase approximation (RPA) this is done by summing
the whole "bubble" series of Fig.\ref{fig:5}, which we recognize as
the familiar plasmon series \cite{pines}. The renormalized diagrams 1
and 2 can now be interpreted as photons interacting with each other
via an exchange of virtual plasmons. The diagram 3 then corresponds to
absorption of two photons with creation of an intermediate plasmon,
plus subsequent decay of the plasmon into the two final photons.

As is well known, exact summation of the plasmon series is equivalent
to replacing the bare polarizability $\Pi^{0}(p_{i})$ by the screened
polarizability
\bqq
\Pi(p_{i}) = {{\Pi^{0}(p_{i})} \over {1 -
U_{0}(\vec{p}_{i}) \Pi^{0}(p_{i})}} \, ,
\eqq
where $$ U_{0}(\vec{q})
= {{4 \pi e^{2}} \over {q^{2}}} $$ is the bare Coulomb interaction.
Our final result for the scattering amplitude then reads
\bqq
M_{fi}^{(2)} = - {{1} \over {\hbar}} ({{e^{2}} \over {2 m c^{2}}})^{2}
\, (4 \pi \hbar c^{2})^{2} \, \sum_{diagrams} (\vec{\epsilon}_{i1} .
\vec{\epsilon}_{i2}) \, (\vec{\epsilon}_{i3} . \vec{\epsilon}_{i4}) \,
{{\Pi^{0}(p_{i})} \over {1 - U_{0}(\vec{p}_{i}) \Pi^{0}(p_{i})}} \, .
\label{eq:m}
\eqq

Because the Lindhard function $\Pi^{0}(p)$ is itself a complicated
function of the 4-momentum transfer $p$, the scattering cross section
resulting from the last expression will also depend in an intricate
way on the frequencies of the incident photons as well as on the
geometry of the situation, which is determined by the wavevectors and
polarizations of all the photons involved. In the rest of this
section, we illustrate just a few particular cases.

First of all, we notice that if the 4-momentum transfer $p_{i}$ is such
that the denominator $1 - U_{0}(\vec{p}_{i})\Pi^{0}(p_{i})$ becomes zero,
the scattering amplitude (\ref{eq:m}) diverges. This corresponds to an
excitation of a real, instead of a virtual, plasmon.
This unphysical divergence, however, is just a
consequence of the random phase approximation, which we have used for
simplicity. Summing a larger class of diagrams and dressing the
plasmon would cause the pole to move away from the real axis, and thus
remove the divergence. Nevertheless, we still expect the scattering
cross section in a realistic metal to be considerably enhanced due to
this "plasmon resonance".

Plasmon resonances can be hit as a function of scattering geometry.
The result (\ref{eq:m}) is valid for a generic geometry. For illustration,
we have investigated simple planar scattering
pictorially described in Fig.\ref{fig:7}. For simplicity, we further
restrict the two incident photons to have equal energies and
perpendicular
wavevectors $\vec{k}_{1} \perp \vec{k}_{2}$ ($\theta = \pi/2$). The
wavevectors of all four photons involved are therefore assumed to lie
in the same plane, the polarization vectors being perpendicular to the
latter. On Fig.\ref{fig:6} we show the dependence of ${{d\sigma} \over
{do}}$ on the angle $\beta$ for $\hbar \Omega_{1} = \hbar \Omega_{2} =
10~eV$. The density of the electron gas was taken to be that of
metallic potassium, with $k_{F}~= 0.73 \times 10^{8}~cm^{-1}$,
corresponding to $E_{F} = 2.03~eV$. The corresponding plasma frequency
is $\hbar\omega_{p} \sim 4~eV$, well below $\hbar \Omega_{1}, \hbar
\Omega_{2}$, as required. One can see sharp peaks corresponding to the
plasmon resonance. The peaks occur for angles $\beta$ such that the
4-momentum of the scattered photon, determined by conservation rules,
exactly fits the requirement for excitation of a real
plasmon in one of the diagrams.

This calculation shows that, at least away from the peaks, the differential
cross section for ultraviolet photons is exceedingly small, of the order of
$10^{-36} cm^{2}$. It is therefore of interest to ask what the
dependence on the {\it photon} frequencies will be, and particularly whether
the cross section could get larger in some other regime. In order
to illustrate the frequency dependence, we consider the particular
case of scattering of two incident photons with opposite wavevectors
(Fig.\ref{fig:7}, $\theta = \pi$),
assuming again all the photons involved to be
polarized perpendicularly to the plane in which the wavevectors lie.
In this case, in the diagram 3 of Fig.\ref{fig:3} the 3--momentum
transfer is zero, and this diagram does not contribute at all. In
diagrams 1 and 2 the energy transfer is zero, and therefore no plasmon
resonance is possible. It is easy now to find the limiting forms of
${{d\sigma} \over {do}}$ for cases $k \ll k_{F}$ (low frequencies) and
$k \gg k_{F}$ (high frequencies).

For the case $k \ll k_{F}$, we expand $\Pi^0(p_{i}) / [1 -
U_{0}(\vec{p}_{i}) \Pi^{0}(p_{i})]$ in (\ref{eq:m}) in powers of $\mid
\vec{p}_{i} \mid$, and keep just the first term, which is of order
$\mid \vec{p}_{i} \mid^2$. The resulting differential cross section is
\bqq
{{d\sigma} \over {do}} = \alpha^2 ({{E_{F}} \over {2 m c^2}})^2
({{k} \over {k_{F}}})^2 k_{F}^{-2} = \alpha^2 ({{E_{F}} \over {2 m
c^2}})^2 {{\Omega^2} \over {c^2 k_{F}^4}} \, . \label{eq:a1}
\eqq
We notice that this result is independent of the angle $\beta$. For
$\hbar \Omega = 10$~eV, it is 24 orders of magnitude larger than the
bare vacuum process (Fig.\ref{fig:12}), whose total integrated cross
section in this regime is known to be \cite{alp}
$$
\sigma = 0.03 \alpha^2 r_{e}^2 ({{\hbar \Omega} \over {m c^{2}}})^6
\, , \,\,\, \hbar \Omega \ll 2 m c^{2}\, ,
$$
where $r_{e} = e^2/m c^2$ is the classical electron radius.

The opposite limit $k \gg k_{F}$ is applicable in the range $\hbar c
k_{F} \ll \hbar \Omega \ll m c^2$. For $k$ not too large, and the
angle $\beta$ not too small, or too close to $\pi$, we expand the
Lindhard function $\Pi^{0}(\vec{p}_{i},\omega_{i})$ in (\ref{eq:m}) in
powers of $1/\mid \vec{p}_{i} \mid$ and keep just the first term,
which is of order $1/\mid \vec{p}_{i} \mid^2$. For the differential
cross section we then obtain
\bqq
{{d\sigma} \over {do}} = {{4} \over
{9 \pi^2}} \alpha^4 {{E_{F}} \over {2 m c^2}} ({{k_{F}} \over {k}})^6
{{1} \over {\sin^4{\beta}}} k_{F}^{-2} = {{4 \alpha^4} \over {9 \pi^2}}
{{E_{F}} \over {2 m c^2}} {{1} \over {\sin^4{\beta}}} {{c^6 k_{F}^4}
\over {\Omega^6}} \label{eq:a2} \, .
\eqq

Fig.\ref{fig:10} shows a log--log plot of ${{d\sigma} \over {do}}$
versus photon frequency $\Omega$ for $\beta = \pi/2$, together with the
asymptotic dependences (\ref{eq:a1}) and (\ref{eq:a2}). We see that
the differential cross section has a maximum for frequencies corresponding
to the photon wavevector $k \sim k_{F}$, where, with the present parameters,
it is of order $10^{-32} cm^2$. This is still a very small value, in fact
somewhat smaller than the maximal photon--photon scattering cross
section in vacuum. However, {\it for photon frequencies in the ultraviolet
and soft X--ray range, up to several keV's, this cross
section is still about 15 orders of magnitude larger than that of the
bare vacuum process} (Fig.\ref{fig:12}). This is what may make it
easier to measure in the end.

So far we have dealt with the 2nd order scattering processes only.
The 3rd order scattering amplitude (\ref{eq:3order}) gives rise to
"triangle" diagrams, like that on Fig.\ref{fig:11}. As we see, in
these diagrams one also encounters the plasmon series, since one of
the vertices corresponds to the interaction term $H_{i2}$. We expect
the contribution of these diagrams to lie somewhere between that of
the 2nd order processes, which we have investigated, and the 4th order
ones, which we will discuss next. We shall not evaluate any of these
3rd order processes explicitly.

\section {\bf Higher order photon--photon scattering processes: the
fourth--order (square) diagrams}

The straightforward second--order calculation of the previous section contains
the main result of this paper, for the photon--photon elastic cross section in
an electron gas. However, it is found, that for $\Omega \gg c k_{F}$, the
cross section has a very fast falloff with frequency, like $\Omega^{-6}$, much
faster than the $\Omega^{-2}$ behaviour known for fourth--order processes
\cite{alp}. In fact, it can be anticipated, that higher order processes should
become increasingly important as frequency increases. Therefore, it is at least
in principle possible that third or fourth--order processes could become
important
when $\Omega$ is both very large with respect to $c k_{F}$, while still very
small relative to $m c^2$ (the optimal photon frequency for vacuum
electron--positron processes). For this reason, we have undertaken the much
heavier task of calculating the fourth--order scattering processes for two
photons in the EG. A second reason for doing this, is that this calculation
will
permit a better formal comparison with photon--photon scattering in vacuum,
which is strictly fourth--order.

We now consider the fourth--order scattering amplitude (\ref{eq:4order}),
which is the non-relativistic analog of the vacuum processes
with the vertex corresponding now to the interaction term $H_{i1}$.

Substituting for $H_{i1}$ from (\ref{eq:hi1}) we get
\begin{eqnarray}
S_{fi}^{(4)} &=& ({{-i} \over {\hbar}})^{4} \, {{1} \over {4!}} \,
({{-i \hbar e} \over {m c}})^{4} \,
\int dx^{4}_{1} \, \ldots  \nonumber \\
& &\int dx^{4}_{4} \,
\langle \Phi_{ph} \!\mid  a_{\vec{k}_{1}^{'}}  a_{\vec{k}_{2}^{'}} \,
T[(\vec{A}(x_{1}).\nabla_{1}) \, \ldots \, (\vec{A}(x_{4}).\nabla_{4})] \,
a_{\vec{k}_{1}}^{\dagger} a_{\vec{k}_{2}}^{\dagger} \mid\! \Phi_{ph}
\rangle \nonumber \\
&\times& \langle \Phi_{FS} \!\mid T[\Psi ^{\dagger}(x_{1}) \Psi (x_{1})
\, \ldots \,
\Psi ^{\dagger}(x_{4}) \Psi (x_{4})] \mid\! \Phi_{FS} \rangle \, ,
\end{eqnarray}
where each of the gradients $\nabla_{i}$ acts only on $\Psi(x_{i})$. Evaluating
this expression in the usual way, using Wick's theorem, we arrive at the
already
mentioned six diagrams (Fig.\ref{fig:1}), corresponding to the six
possibilities for contraction of the $\Psi$ operators.

Making use of
\bqq
\nabla_{\vec{x}} G^{0}(x,y) = (2 \pi)^{-4} \int d^{4}p \, (i \vec{p})
e^{i p (x-y)} G^{0}(p) \, ,
\eqq
and of relations (\ref{eq:s}) and (\ref{eq:t}), and collecting together all the
numerical factors we get finally, omitting all details, an expression for
$M_{fi}^{(4)}$
\bqq
M_{fi}^{(4)} = {{2 i \hbar^2} \over {\pi^2}} ({{e} \over {m}})^{4}
\sum_{diagrams} \, I_{i} \, .
\eqq
We have denoted as $I_{i}$ the integral corresponding to the i-th diagram
\bqq
I_{i} = \int d^{4}p \, (\vec{\epsilon}_{i1} . \vec{p}_{i1}) \, \ldots \,
(\vec{\epsilon}_{i4} . \vec{p}_{i4}) G^{0}(p_{i1}) G^{0}(p_{i2}) G^{0}(p_{i3})
G^{0}(p_{i4}) \, , \label{eq:4integral}
\eqq
where $\vec{\epsilon}_{in}$ is the photon polarization unit vector in $i$-th
diagram, vertex $n$.

Now since each electron Green's function $G^{0}$ consists of two parts
corresponding to electrons and holes, respectively, each of the above integrals
splits into 16 terms. Each of these will contain a product of 4 step
functions $\Theta(\pm(\mid \vec{p_{j}} \mid - k_{F}))$, which defines an
integration region that is an intersection of interiors and/or exteriors of
four mutually displaced Fermi spheres. For general direction and magnitudes of
photon wavevectors, the integration over such region would be quite intricate.
For the sake of simplicity, we have not attempted to evaluate all integrals for
a general situation. In the Appendix we present details of a particular
calculation for the special geometry of {\it forward scattering} of two photons
{\it with opposite wavevectors} (i.e. two initial photons with wavevectors
$\vec{k}, \, -\vec{k}$ scatter into the same two final photons), where the
integrations turn out to be particularly easy. We shall just quote here the
final results of this rather tedious calculation, where approximations have
also been made based on relations (\ref{eq:nonrel1}) and (\ref{eq:nonrel2}).

We have considered two limiting cases for the photon wavevector. First we take
the case $k \ll k_{F}$, which applies very well for ultraviolet photon energies
$\hbar \Omega \ll$ 1~keV.
We have obtained the result, Eq. (\ref{eq:sumint1}), which, after passing from
the rescaled variables (used throughout the appendix) back to the original
ones,
reads
\bqq
\sum_{diagrams} I_{i} = ({{2 m} \over {\hbar}})^{3} k_{F}
{{8 \pi i} \over {3}} B(\phi_{1},\phi_{2},\phi_{1}^{'},\phi_{2}^{'})
({{k} \over {k_{F}}})^3
({{\hbar k_{F}^{2}} \over {2 m}})^4 {{1} \over {\Omega^4}} \, ,
\eqq
where $k$ and $\Omega$ are the wavevector and the frequency of the photon. The
function $B(\phi_{1},\phi_{2},\phi_{1}^{'},\phi_{2}^{'})$ is a geometrical
factor, which accounts for the polarization of the photons, and is defined in
equation (\ref{eq:gb}) in the Appendix.

Substituting for $\Omega$ the bare photon frequency $\Omega = c k$, we get for
$M_{fi}^{(4)}$
\bqq
M_{fi}^{(4)} = - {{256 B} \over {3 \pi}} {\alpha}^2 ({{E_{F}} \over
{2 m c^{2}}})^{3/2} {{c^{2} \hbar^{-2}} \over {\Omega}} E_{F}^2 k_{F}^{-1} \, .
\eqq
The corresponding fourth--order differential scattering cross section
${{d\sigma} \over {do}}$ is then finally, according to (\ref{eq:dsdo})
\bqq
{{d\sigma} \over {do}} = {{32 B^{2}} \over {9 \pi^{4}}} \alpha^{4}
({{E_{F}} \over {m c^{2}}})^{3} ({{E_{F}} \over {\hbar\Omega}})^{4} k_{F}^{-2}
\, , \,\,\, \omega_{P} < \sim \Omega \ll c k_{F} \, .  \label{eq:4reslk}
\eqq
To get an order of magnitude for ${{d\sigma} \over {do}}$ we
substitute
in the last expression the values typical for metals, i.e. $k_{F} \sim 10^{8}
cm^{-1}$ and $E_{F} \sim 5$~eV.  With $\hbar\Omega \sim E_{F}$, we obtain
\bqq
{{d\sigma} \over {do}} \sim (4 \times 10^{-2}) \, . \, (3 \times 10^{-9}) \, .
(10^{-5})^{3} \, . \, 1 \, . \, (10^{8})^{-2}
\sim 10^{-41} cm^{2} \, . \label{eq:estim} \nonumber
\eqq

The second limiting case we have investigated is the case of photon wavevectors
falling above $2 k_{F}$, which means X-ray photons with
energies $\hbar \Omega > \sim$ 1~keV. The sum over the diagrams is given by
(\ref{eq:iss}), which translated back to true variables reads
\bqq
\sum_{diagrams} I_{i} = ({{2 m} \over {\hbar}})^{3} k_{F}
{{- 64 \pi i} \over {15}} B(\phi_{1},\phi_{2},\phi_{1}^{'},\phi_{2}^{'})
({{k} \over {k_{F}}})^4
({{\hbar k_{F}^{2}} \over {2 m}})^4 {{1} \over {\Omega^4}} \, .
\eqq
Substituting again the bare photon frequency for $\Omega$ we obtain the
scattering amplitude
\bqq
M_{fi}^{(4)} = {{512 B} \over {15 \pi}} {\alpha}^2 ({{E_{F}} \over {m
c^{2}}})^2
c^{2} \hbar^{-1} E_{F} k_{F}^{-1} \, ,
\eqq
and we notice that this does not depend on $\Omega$. The differential
scattering cross section then equals
\bqq
{{d\sigma} \over {do}} = ({{32 B} \over {15 \pi^2}})^2 \alpha^{4}
({{E_{F}} \over {m c^{2}}})^{4} ({{E_{F}} \over {\hbar\Omega}})^{2}
k_{F}^{-2} \, , \,\,\, 2 \hbar c k_{F} < \hbar \Omega \ll m c^2  \, .
\label{eq:4resgk}
\eqq
We estimate the order of magnitude of ${{d\sigma} \over {do}}$ for
$k \sim 2 k_{F}$ and obtain
\bqq
{{d\sigma} \over {do}} \sim (4 \times 10^{-2}) \, . \, (3 \times 10^{-9}) \, .
\, (10^{-5})^4 \, . \, (10^{-3})^2 \, . \, (10^{8} cm^{-1})^{-2}
\sim 10^{-52} cm^2 \, . \nonumber
\eqq

Equations (\ref{eq:4reslk}) and (\ref{eq:4resgk}) represent the main results of
this section. We can now compare them with their equivalent results in vacuum,
namely \cite{alp}
\begin{eqnarray}
\sigma = 0.03 \, \alpha^2 r_{e}^2 ({{\hbar \Omega} \over {m c^{2}}})^6 \, ,
\,\,\, \hbar \Omega \ll 2 m c^{2} \label{eq:vac1} \\
\sigma = 4.7 \, \alpha^4 ({{c} \over {\Omega}})^2 \, , \,\,\,
\hbar \Omega \gg 2 m c^{2} \, ,
\end{eqnarray}
where $\sigma$ is the total integrated cross section for unpolarized photons
and
$r_{e} = e^2/m c^2$ is the classical electron radius. We note, first of
all, that the $\alpha^2$ dependence at low frequencies is the same as in our
low frequency second--order result of Section 3, Eq.(\ref{eq:a1}), while the
high frequency $\alpha^4$ behaviour is the same as in both the second--order
result
of Eq.(\ref{eq:a2}), and in the present fourth--order results (\ref{eq:4reslk})
and (\ref{eq:4resgk}). Secondly, we see that the $\Omega^{-2}$ frequency
dependence (assuming $\Omega_{1} = \Omega_{2} = \Omega$) is the same for both
4th order
in vacuum and 4th order in the EG, provided the high-frequency regime is
reached
in each case. Numerically, however, the coefficient in front of $\Omega^{-2}$
is much smaller in the EG case. At lower frequencies, instead, we find a new
regime in the EG case, where the 4th order cross section falls off like
$\Omega^{-4}$, Eq.(\ref{eq:4reslk}). This regime does not exist in vacuum, and
is clearly due to the presence of a Fermi surface in the electron gas problem.

We can now try an overall graphical comparison of all these results. This is
sketched in Fig.\ref{fig:12}. We see that the 2nd order processes in the EG are
dominant for photon energies in the range from $\omega_{P}$ up to
$\sim 10^4~\div~10^5$~eV. For X--ray photons of $10^5$~eV the relativistic
vacuum processes become important and dominate at all
higher photon frequencies. Hence, the 4th order EG contributions are always
negligible, and by a huge factor, in spite of their slower falloff with photon
energy. It is reasonable, therefore, to assume without proof that similar
conclusions will apply to the third order processes, which we have accordingly
ignored.

\section {\bf Discussion, and possible experimental verification.}

We have considered the problem of elastic ultraviolet and X--ray photon-photon
scattering inside a solid, idealized as a free electron gas. We find that the
presence of the electron gas should give rise to new scattering processes, much
more important than those present in vacuum. There, only 4th order processes,
important for $\gamma$--ray photons in the MeV range, are operative. In the
nonrelativistic electron gas, instead, the existence of two-photon vertices
introduces large second order processes, which are most efficient for photon
wavevector roughly equal to the Fermi wavevector of the metal. We also predict
important plasmon resonances to take place, and an angular dependence different
from that of the vacuum processes.

We have also recalculated the 4th order processes in the EG. We find that these
are themselves different from those in vacuum, mostly due to the breakdown of
electron-hole symmetry, and their importance increases as the photon frequency
decreases. However, the plasma frequency $\omega_{P}$ represents in practice a
lower bound for the frequency of a photon, if it should penetrate inside a
metal, and we have found the 4th order processes to be negligible for
frequencies higher than $\omega_{P}$. The overall situation is summarized by
Fig.\ref{fig:12}, which also gives an order of magnitude of cross sections for
a metal such as potassium, chosen because of its low plasma frequency.

Let us now briefly consider the possibilities for experimental verification of
our calculated electron-mediated photon-photon elastic scattering processes.
We shall discuss two cases, both extremely idealized.
In the first we consider the use of a powerful pulsed laser
source, operating in the ultraviolet region, $\hbar \Omega \sim$~10~eV. The
second case will be that of an X--ray in 1 keV range, such as that, which can
be obtained from a synchrotron source (plus undulators).

In order to estimate the scattering cross section for the first case, we refer
to Fig.\ref{fig:6}. Making use of a plasmon resonance, we can expect
${{d\sigma} \over {do}} \sim 10^{-35}$~cm$^2$ for a metal. One could
think, for example, of a crossed-beam experiment. Two laser beams are focused
on a small area and penetrate into the metal and the scattered radiation is
picked up by a detector. Without worrying about problems of power
dissipation, let us think of one laser pulse per second, of a duration
of $T$~picoseconds and a peak power of $I \times 10^{10}$~Watts. The pulse
contains an energy of $\sim~I~T~\times~10^{-2}$~J, which
corresponds to $\sim~I~T~\times~10^{16}$~photons with energy 10~eV. The spatial
length of the pulse is $3~T~\times~10^{-2}$~cm $=~3~T~\times~10^6$~\AA.
Imagining a penetration depth
of $\sim 200$ \AA, the number of photons inside the metal is
$I~\times~10^{12}$.
The total scattering cross section, as seen by another photon, is therefore
$\sim~I~\times~10^{-23}$~cm$^2$. If the second beam, having the same parameters
as the first one, is focused on the area of (3~$\mu$m)$^2~\sim~10^{-7}$~cm$^2$,
the number of collisions during the pulse is ${{I \times 10^{-23} cm^2} \over
{10^{-7} cm^2}}~\times~I~\times~T~\times~10^{16}~\sim~I^2~T$. For a peak
power of, e.g., $10^{10}$~W ($I = 1$) and a pulse duration of 1~ps ($T = 1$)
we can expect about one scattered photon per unit solid angle per second.
This flux should be, presumably, detectable with present experimental
techniques.

In the case of a synchrotron source, one can work in the soft X--ray region,
where the 2nd order scattering cross section has a maximum. According to
Fig.\ref{fig:10}, this occurs for photon energies
$\hbar \Omega~\sim~\hbar c k_{F}$ in the keV range, where one has
${{d\sigma} \over {do}}~\sim~10^{-32}$~cm$^2$.
The penetration depth in the metal in this region might be
$\sim~10^4$~\AA. To be more specific, we consider a
beam of 1 keV radiation, consisting of $10^9$ pulses per second,
with pulse duration 100~ps and peak photon flux $10^{20}$~photon~s$^{-1}$. If
the cross section of the beam is taken $\sim~10^{-7}$~cm$^2$, the resulting
number of scattered photons is again $\sim$~1~photon per second and unit solid
angle, which is similar to the crossed UV laser case, and should have no sample
burning problems.

In all cases, it is clear that, apart from the products of photon-photon
collisions (which we are after) there will be a background of photons scattered
by other entities due to different mechanisms. There will be, for example,
single-photon processes, like photon scattering from free electrons. The order
of magnitude of the corresponding cross section can be estimated from classical
Thomson formula to be of order $d\sigma \sim r_{e}^{2} \sim 10^{-25} cm^{2}$,
and is in general larger than the photon-photon cross section, at least for
reasonable intensities. These single-photon processes, however, should be
distinguishable from the two-photon processes of our interest, either by
requiring photon-photon coincidence, or by making use of either the quadratic
intensity dependence of the photon-photon events, or of their polarization
dependence (\ref{eq:m}).

A subtler complication is that of additional contribution of interband
transitions to two-photon processes. Since interband transitions are vertical,
the intermediate states in the scattering process can be real ones, at
particular
photon frequencies. Care should be taken, therefore, to stay away from these
frequencies as much as possible. As for inelastic processes, which probably
represent a big problem, one may perhaps think of using the typical elastic
polarization dependence obtained in Eq. (\ref{eq:m}), to subtract them out.

Yet another possible route, although admittedly very speculative, could be
studying the subtle changes in the interference pattern which one can expect to
take place at very high photon intensity in a classical interference
experiment.
The classical theory of interference
\cite{loudon} is of course based on the superposition principle, i.e. on
photon-photon scattering being {\it exactly} zero. If that is no longer true,
one can expect that correlation effects building up between the photons should
in principle change intensities in the interference pattern. De Martini et al.
\cite{demart}, for example, have recently shown how random selection of
polarization in front of the interferometer presents precisely a realization of
this kind of effect. In that case, when the beams are very intense, photons
become correlated with one another through the random polarizer. As a result,
their distribution
tends to become Bose-like, which can be imagined as "trains" of photons going
down one or the other slit separately. This has been shown \cite{demart} to
weaken the interference pattern in a characteristic and measurable manner.

In principle, when the beam is sufficiently intense and the "slits" are
metallic
(e.g., a half-silvered mirror), a similar kind of modification of the
interference
pattern should be expected even {\it without} the random polarizer. Our
estimates of the cross section could in principle be used to evaluate the
threshold value of the intensity in a well-defined interference geometry.

A general remark may be in order, before closing this paper. As things stand,
there are two elements which suggest that our calculation might
remain, at least for while, of purely academic interest. The first
element is that our calculated photon-photon cross sections are still very
small. Detection of these processes may require a nontrivial experimental
effort. The second point is that detection of our proposed process does not in
itself provide new basic information, on either the photon, or the system.

Our viewpoint on these issues is open. Without embarking in a discussion of why
one should or should not try to measure the process we calculated, we have
simply meant to point out, for future record, that there are new photon-photon
processes, which were not discussed so far, and which are in principle of
measurable intensity.

\nonum
\section {\bf Appendix: details of fourth--order calculation.}

To avoid problems with ill-defined expressions we shall evaluate
$M_{fi}^{(4)}$ for the case of scattering at a small angle, i.e.
$\vec{k}_{1}=\vec{k}$, $\vec{k}_{2}=-\vec{k}$, $\vec{k}_{1}^{'}=
\vec{k}+\vec{q}$, $\vec{k}_{2}^{'} = - \vec{k}-\vec{q}$, where $\vec{q}$ must
be such that
$\mid \vec{k}+\vec{q} \mid = \mid \vec{k} \mid$, and then perform the limit
$\mid \vec{q} \mid \rightarrow 0$ to obtain the forward scattering amplitude.
Because of the transversality of the vector potential, in the limit
$\mid \vec{q} \mid \rightarrow 0$ we have
\bqq
\vec{k} . \vec{e}_{1} = \vec{k} . \vec{e}_{2}
= \vec{k} . \vec{e}_{1}^{'} = \vec{k} . \vec{e}_{2}^{'} = 0 \, . \label{eq:pol}
\eqq
In order to deal with dimensionless variables we rescale quantities in the
integrals (\ref{eq:4integral}) and introduce dimensionless frequencies and
wavevectors by the relations
\begin{eqnarray}
\omega  =  {{\hbar k_{F}^{2}} \over {2 m}} \omega^{'} \, , \\
k =  k_{F} k^{'} \, .
\end{eqnarray}
The integrals $I_{i}^{'}$ calculated with the rescaled variables
$\omega^{'}$, $k^{'}$ are related to the original ones by
\bqq
I_{i} = ({{2 m} \over {\hbar}})^{3} k_{F} I_{i}^{'} \, ,
\eqq
and in the following the primes in all dimensionless quantities will be
omitted.

As already mentioned, each of the integrals (\ref{eq:4integral}) splits
into 16 terms corresponding to all possible combinations of electrons and holes
participating in the process. Some of these terms are, however, immediately
seen to be equal to zero. Apart from obvious vanishing of the
terms in which all 4 internal lines correspond simultaneously to electrons,
resp. holes, terms in diagrams 1 and 3 in which there is an electron and a hole
in the state with the same 3-momentum, namely $\vec{p}_{2} = \vec{p}_{4}$, have
to vanish, too, due to the vanishing of the integration region.
In the Tab.~\ref{t1} we enumerate all possibly non-zero contributions to
integrals $I_{1}$, $I_{2}$, $I_{3}$, corresponding respectively to the diagrams
1,2,3 (Fig.\ref{fig:8}), where $+$ sign denotes a hole and $-$ sign an
electron.

Each of the terms has a general form
\bqq
\int d^{3}\vec{p} \, \int d\omega \, (\vec{\epsilon}_{i1} . \vec{p}_{i1}) \,
\ldots \, (\vec{\epsilon}_{i4} . \vec{p}_{i4}) \,
{{\Theta[\pm(\mid \vec{p}_{i1} \mid - 1)]} \over
{\omega_{i1} - \vec{p}_{i1}^2 \pm i \eta}} \,
\ldots \,
{{\Theta[\pm(\mid \vec{p}_{i4} \mid - 1)]} \over
{\omega_{i4} - \vec{p}_{i4}^2 \pm i \eta}} \, ,
\eqq
where the momenta $\vec{p}_{i1},\ldots,\vec{p}_{i4}$ depend on $\vec{p}$
and frequencies $\omega_{i1},\ldots,\omega_{i4}$ depend on $\omega$.
Performing the frequency integrals using the residue theorem and passing to the
limit $\mid \vec{q} \mid \rightarrow 0$ wherever this can be done in a
straightforward way we get the following expressions
\begin{eqnarray}
I_{1a} + I_{1b}  &=&  4 \pi i \int d^{3}\vec{p} [\ldots] \Theta(1- \mid \vec{p}
\mid) \Theta(\mid \vec{p} - \vec{k} \mid -1) \nonumber \\
&\times& {{-4 \vec{k} . \vec{p} + 2 k^{2}} \over
{(\Omega + 2 \vec{k} . \vec{p} - k^{2})^{2}
 (\Omega - 2 \vec{k} . \vec{p} + k^{2})^{2}}} \\
I_{2a} + I_{2b}  &=&  4 \pi i \int d^{3}\vec{p} [\ldots] \,
\Theta(1 - \mid \vec{p} \mid)
\Theta(1 - \mid \vec{p} + \vec{k} \mid) \Theta(\mid \vec{p} - \vec{k} \mid - 1)
\nonumber \\
&\times& {{1} \over {4 \vec{k} . \vec{p}}} \,
{{1} \over {(\Omega - 2 \vec{k} . \vec{p} + k^{2})^{2}}} \\
I_{3a} + I_{3b}  &=&  4 \pi i \int d^{3}\vec{p} [\ldots] \,
\Theta(1 - \mid \vec{p} \mid)
\Theta(1 - \mid \vec{p} + \vec{k} \mid) \Theta(\mid \vec{p} - \vec{k} \mid - 1)
\nonumber \\
&\times& {{1} \over {4 \vec{k} . \vec{p}}} \,
{{1} \over {\Omega - 2 \vec{k} . \vec{p} + k^{2}}} \,
{{1} \over {-\Omega - 2 \vec{k} . \vec{p} + k^{2}}} \\
I_{2d} + I_{2e}  &=&  4 \pi i \int d^{3}\vec{p} [\ldots] \,
\Theta(\mid \vec{p} \mid - 1)
\Theta(1 - \mid \vec{p} + \vec{k} \mid) \Theta(\mid \vec{p} - \vec{k} \mid - 1)
\nonumber \\
&\times& {{1} \over {4 \vec{k} . \vec{p}}} \,
{{1} \over {(\Omega + 2 \vec{k} . \vec{p} + k^{2})^{2}}} \\
I_{3d} + I_{3e}  &=&  4 \pi i \int d^{3}\vec{p} [\ldots] \,
\Theta(\mid \vec{p} \mid - 1)
\Theta(1 - \mid \vec{p} + \vec{k} \mid) \Theta(\mid \vec{p} - \vec{k} \mid - 1)
\nonumber \\
&\times& {{1} \over {4 \vec{k} . \vec{p}}} \,
{{1} \over {\Omega + 2 \vec{k} . \vec{p} + k^{2}}} \,
{{1} \over {-\Omega + 2 \vec{k} . \vec{p} + k^{2}}}
\end{eqnarray}
\begin{eqnarray}
I_{2c}  &=&  -2 \pi i \int d^{3}\vec{p} [\ldots] \,
\Theta(1 - \mid \vec{p} \mid)
\Theta(\mid \vec{p} + \vec{k} \mid - 1) \Theta(\mid \vec{p} - \vec{k} \mid - 1)
\nonumber \\
&\times& {{1} \over {4 \vec{k} . \vec{p}}} \,
\left[ {{1} \over {(\Omega + 2 \vec{k} . \vec{p} + k^{2})^{2}}} -
  {{1} \over {(\Omega - 2 \vec{k} . \vec{p} + k^{2})^{2}}} \right] \\
I_{3c}  &=&  -2 \pi i \int d^{3}\vec{p} [\ldots] \,
\Theta(1 - \mid \vec{p} \mid)
\Theta(\mid \vec{p} + \vec{k} \mid - 1) \Theta(\mid \vec{p} - \vec{k} \mid - 1)
\nonumber \\
&\times& \, {{1} \over {2 \Omega}} \,
\left[ {{1} \over {\Omega + 2 \vec{k} . \vec{p} + k^{2}}} \,
  {{1} \over {\Omega - 2 \vec{k} . \vec{p} + k^{2}}} -
  {{1} \over {\Omega - 2 \vec{k} . \vec{p} - k^{2}}} \,
  {{1} \over {\Omega + 2 \vec{k} . \vec{p} - k^{2}}} \right] \, ,
\end{eqnarray}
where $\Omega$ is the photon frequency.
We have used the relations~(\ref{eq:pol}) and introduced the notation
\bqq
[\ldots] = (\vec{\epsilon}_{i1} . \vec{p}) \, (\vec{\epsilon}_{i2} . \vec{p})
\, (\vec{\epsilon}_{i3} . \vec{p}) \, (\vec{\epsilon}_{i4} . \vec{p}) =
(\vec{e}_{1} . \vec{p}) \, (\vec{e}_{2} . \vec{p}) \,
(\vec{e}_{1}^{'} . \vec{p}) \, (\vec{e}_{2}^{'} . \vec{p})
\eqq
in the above expressions. In several terms we have
performed a shift and/or an inversion of the integration
variable $\vec{p}$. Terms 2f and 3f are identically zero since the
corresponding
$\Theta$ functions have zero product.

Terms $I_{1c}$ - $I_{1f}$ and $I_{2g}$ - $I_{2n}$ require a comment. They
contain products like $\Theta(1 - \mid \vec{p} \mid) \,
\Theta(\mid \vec{p} + \vec{q} \mid - 1)$, which restrict the integration volume
to a thin shell about a half of the surface of the Fermi sphere
(see Fig.\ref{fig:9}).
The volume element of the integration region can then be written as
$$ dV = \pm d\vec{S}.\vec{q} = \pm dS \vec{p}.\vec{q} \, . $$
At the same time these terms contain one denominator of the form
$$ (\vec{p} +\vec{q})^{2} - \vec{p}^{2} = 2 \vec{p}.\vec{q} + \vec{q}^2
\rightarrow 2 \vec{p}.\vec{q} \, ,$$
which in the limit $\mid \vec{q} \mid \rightarrow 0$ precisely cancels the
same product in the volume element.
In the limit $\mid \vec{q} \mid \rightarrow 0$ therefore these terms become
well defined surface integrals. Summing all the contributions from the
diagrams 1, resp. 2, we obtain expressions
\begin{eqnarray}
I_{1s} &=&  I_{1c} + I_{1d} + I_{1e} + I_{1f} \nonumber \\
       &=&  - \pi i \int_{\mid\ \vec{p} \mid = 1} dS [\ldots]
{{1} \over {\Omega + 2 \vec{k} . \vec{p} - k^{2}}} \,
{{1} \over {-\Omega + 2 \vec{k} . \vec{p} - k^{2}}} \\
I_{2s} &=&  I_{2h} + I_{2i} + I_{2j} + I_{2l} + I_{2m} + I_{2n} \nonumber \\
       &=&  - \pi i \int_{\mid\ \vec{p} \mid = 1} dS [\ldots]
{{1} \over {\Omega + 2 \vec{k} . \vec{p} + k^{2}}} \,
{{1} \over {\Omega - 2 \vec{k} . \vec{p} + k^{2}}} \, ,
\end{eqnarray}
where the integration region is the whole surface of the Fermi sphere. Terms
$I_{2g}$ and $I_{2k}$ are zero again because of vanishing of the corresponding
product of $\Theta$ functions.

Now we must take into account also contributions of other 3 diagrams which
differ from those on Fig.\ref{fig:8} by the orientation of the internal loop.
Due to the
absence of charge-conjugation symmetry, Furry's theorem does not apply, and the
contributions of diagrams which differ from each other by the orientation of
the internal loop may not be equal. It can be easily seen
that the contributions of diagrams 1 and 3 do not depend on the orientation,
while in case of diagram 2 the change of the orientation is equivalent to the
change of sign of the photon frequency $\Omega$. Therefore the sum of
contributions of all the diagrams can be written as
\begin{eqnarray}
\sum_{all \, diagrams} I_{i} &=& 2(I_{1a} + I_{1b}) + 2I_{1s} + 2(I_{3a} +
I_{3b}) + 2I_{3c} + 2(I_{3d} + I_{3e}) \nonumber\\
&+& I_{2a}(\Omega) + I_{2a}(-\Omega) + I_{2b}(\Omega)
+ I_{2b}(-\Omega) + I_{2c}(\Omega) + I_{2c}(-\Omega) \nonumber\\
&+& I_{2d}(\Omega) + I_{2d}(-\Omega)+ I_{2e}(\Omega)
+ I_{2e}(-\Omega) + I_{2s}(\Omega) + I_{2s}(-\Omega) \, . \label{eq:alldiag}
\end{eqnarray}

At this point we shall perform some approximations. First we make use of the
conditions (\ref{eq:nonrel1}), (\ref{eq:nonrel2}), which we assumed at the very
beginning of our non-relativistic treatment. We write the bare photon
dispersion relation $\Omega = c k$ in the rescaled variables as
\bqq
\Omega = A k \, ,
\eqq
where
\bqq
A = {{2mc} \over {\hbar k_{F}}} \, .
\eqq
Substituting to this expression a typical Fermi wavevector of metals
$k_{F}~\sim~10^8$~cm$^{-1}$ we observe that
\bqq
A \sim 10^{2} \gg 1  \label{eq:Agg1}
\eqq
holds for $A$. Actually, the last condition is just the square root of the
condition (\ref{eq:nonrel2}). Moreover, the condition (\ref{eq:nonrel1})
now turns out to be equivalent to
\bqq
k \ll A \, . \label{eq:klla}
\eqq

Because there is always a hole participating in the process, the integration
variable $\vec{p}$ must fall {\it inside} at least one of the 3 Fermi spheres
defined by $\mid \vec{p} \mid \leq 1$, $\mid \vec{p} + \vec{k} \mid \leq 1$ and
$\mid \vec{p} - \vec{k} \mid \leq 1$. This implies
\bqq
\mid \vec{p} \mid \leq 1 + k \, ,
\eqq
and due to (\ref{eq:Agg1}) and (\ref{eq:klla}) also
\bqq
\mid \vec{p} \mid \ll A \, . \label{eq:plla}
\eqq

Now in the denominators of our integrals we encounter terms like
$\Omega \pm 2 \vec{k}.\vec{p} \pm k^{2} = A k \pm 2 \vec{k}.\vec{p} \pm k^{2}$,
and the relations (\ref{eq:klla}), (\ref{eq:plla}) and (\ref{eq:Agg1})
immediately suggest that we can expand the integrands in powers of $1/A$ and
keep just the first non--zero term. Since the expression
(\ref{eq:alldiag}) is an even function of $\Omega$ and therefore of $A$, the
expansion contains only even powers, and represents actually an expansion in
powers of ${{1} \over {A^{2}}} = {{E_{F}} \over {2 m c^{2}}}$. The first
non--zero term turns out to be of the order $1/A^4$ in all the integrands, and
neglecting terms of the order of $1/A^6$ we obtain respectively
\bqq
2(I_{1a} + I_{1b})  =  \pi i \int d^{3}\vec{p} [\ldots] \Theta(1- \mid \vec{p}
\mid) \Theta(\mid \vec{p} - \vec{k} \mid - 1) { {16 (k^2 - 2 \vec{k} .
\vec{p})}
\over {\Omega^4} }
\eqq
\begin{eqnarray}
& & 2(I_{3a} + I_{3b}) + (I_{2a} + I_{2b})(\Omega) + (I_{2a} + I_{2b})(-\Omega)
= \nonumber \\
& & \pi i \int d^{3}\vec{p} [\ldots] \, \Theta(1 - \mid \vec{p} \mid)
\Theta(1 - \mid \vec{p} + \vec{k} \mid) \Theta(\mid \vec{p} - \vec{k} \mid - 1)
{{4} \over {\vec{k} . \vec{p}}} \,
{ {(2 \vec{k} . \vec{p} - k^2)^2} \over {\Omega^4} } \\
& & 2(I_{3d} + I_{3e}) + (I_{2d} + I_{2e})(\Omega) + (I_{2d} + I_{2e})(-\Omega)
= \nonumber \\
& & \pi i \int d^{3}\vec{p} [\ldots] \, \Theta(\mid \vec{p} \mid - 1)
\Theta(1 - \mid \vec{p} + \vec{k} \mid) \Theta(\mid \vec{p} - \vec{k} \mid - 1)
{{4} \over {\vec{k} . \vec{p}}} \,
{ {(2 \vec{k} . \vec{p} + k^2)^2} \over {\Omega^4} }
\end{eqnarray}
\begin{eqnarray}
I_{C} &=& 2 I_{3c} + I_{2c}(\Omega) + I_{2c}(-\Omega) \nonumber \\
&=& \pi i \int d^{3}\vec{p} [\ldots] \, \Theta(1 - \mid \vec{p} \mid)
\Theta(\mid \vec{p} + \vec{k} \mid - 1) \Theta(\mid \vec{p} - \vec{k} \mid - 1)
{ {- 16 k^2} \over {\Omega^4} } \, , \label{eq:ic} \\
I_{S} &=& 2 I_{1s} + I_{2s}(\Omega) + I_{2s}(-\Omega) \nonumber \\
&=& \pi i \int_{\mid\ \vec{p} \mid = 1} dS [\ldots]
{ {- 4 k^2} \over {\Omega^4} } (2 \vec{k} . \vec{p} + k^2) \, , \label{eq:is}
\end{eqnarray}
where we have already grouped together some terms.

Now we could, in principle, calculate the above integrals in a straightforward
way and obtain the scattering amplitude for any $k$, restricted only by the
condition (\ref{eq:klla}). For sake of simplicity we shall do the integrations
explicitly for 2 limiting cases, in which the integration region becomes
particularly simple, first for $k \ll 1$, and then for $2 < k \ll A$.

The condition $k \ll 1$ allows us once again to pass in several terms to
surface integration (in the same way as above when we were performing the limit
$\mid \vec{q} \mid \rightarrow 0$). In the limit $k \ll 1$, the
integration regions of the terms 1a,1b,2a,2b,3a,3b,2d,2e,3d and 3e become the
same and equal to the half of the surface of the Fermi sphere \\
$\mid \vec{p} \mid = 1$, for which $\vec{k}.\vec{p} \leq 0$ holds. Collecting
these terms together and denoting their sum as $I_{I}$ we obtain
\begin{eqnarray}
I_{I} &=& 2(I_{1a} + I_{1b}) + 2(I_{3a} + I_{3b}) + (I_{2a} + I_{2b})(\Omega) +
(I_{2a} + I_{2b})(-\Omega) \nonumber\\
&+& 2(I_{3d} + I_{3e}) + (I_{2d} + I_{2e})(\Omega) + (I_{2d} + I_{2e})(-\Omega)
\nonumber \\
&=& {{-\pi i} \over {\Omega^{4}}} \int_{\vec{k}.\vec{p} \leq 0} dS [\ldots] \,
(8 k^{4} + 16 k^{2} \vec{k}.\vec{p}) \, . \label{eq:ii}
\end{eqnarray}
The remaining terms in (\ref{eq:alldiag}) are the volume integral $I_{C}$
(\ref{eq:ic}) and the surface integral $I_{S}$ (\ref{eq:is}).

Since we are considering the limiting case $k \ll 1$, it is now enough to
identify in the sum $I_{I} + I_{C} + I_{S}$ terms which contain the leading
power of $k$. In $I_{I}$ the leading term in $k$ is of order $k^{3}$. It is
easily seen that the integration volume of the integral $I_{C}$ behaves as
$k^{3}$ and the contribution of this term is therefore of order $k^{5}$.
In the surface integral $I_{S}$ we encounter a term
$\sim k^{2} \vec{k}.\vec{p}$, which, however, vanishes after the integration
over the whole surface of the Fermi sphere and the remaining term contributes
as
$k^{4}$. In the limit $k \ll 1$ therefore the dominant contribution to the
scattering amplitude comes from the leading term in (\ref{eq:ii}), which is
readily evaluated as
\bqq
\sum_{all \, diagrams} I_{i} = {{8 \pi i} \over {3}} {{k^{3}} \over
{\Omega^{4}}} B(\phi_{1},\phi_{2},\phi_{1}^{'},\phi_{2}^{'}) \, .
\label{eq:sumint1}
\eqq
In the last expression we introduced a function
\begin{eqnarray}
& & B(\phi_{1},\phi_{2},\phi_{1}^{'},\phi_{2}^{'}) =
\int_{0}^{2\pi} d\phi \cos(\phi-\phi_{1}) \cos(\phi-\phi_{2})
\cos(\phi-\phi_{1}^{'}) \cos(\phi-\phi_{2}^{'}) = {{\pi} \over 4} \times
\nonumber \\
&\times& \left[ \cos (\phi_{1} + \phi_{2} - \phi_{1}^{'} - \phi_{2}^{'})
+ \cos (\phi_{1} - \phi_{2} + \phi_{1}^{'} - \phi_{2}^{'}) +
\cos (\phi_{1} - \phi_{2} - \phi_{1}^{'} + \phi_{2}^{'}) \right] ,
\label{eq:gb}
\end{eqnarray}
which represents a factor taking into account the polarization of the photons,
and $\phi_{1},\phi_{2},\phi_{1}^{'},\phi_{2}^{'}$ are direction angles of the
unit polarization vectors of the photons.

In another limiting case, $2 < k \ll A$, the Fermi spheres are displaced from
one another by a distance larger than 2 radii, and therefore do not intersect
anymore. The integration region of integrals $I_{2a} + I_{2b}$ and
$I_{3a} + I_{3b}$ becomes empty and that of the remaining volume integrals
becomes the whole interior of one of the spheres. Denoting the sum of all
volume integrals as $I_{V}$ we obtain
\begin{eqnarray}
& & I_{V} = {{\pi i} \over {\Omega^{4}}} \int d^{3}\vec{p} [\ldots] \,
\Theta(1 - \mid \vec{p} \mid) \left[ -32 \vec{k} . \vec{p} +
{{4 (2 \vec{k} . \vec{p} - k^2)^2} \over {\vec{k} . (\vec{p} - \vec{k})}}
\right] = {{4 \pi i} \over {\Omega^{4}}} \times \nonumber \\
&\times&  B(\phi_{1},\phi_{2},\phi_{1}^{'},\phi_{2}^{'})
\left\{ {{k^4} \over {45}} (-33 + 40 k^2 - 15 k^4) +
{{1} \over {6}} (k^3 - 3 k^5 + 3 k^7 - k^9) \ln {{k - 1} \over {k + 1}}
\right\} . \label{eq:iv}
\end{eqnarray}
Apart from this, there is also the surface integral $I_{S}$ (\ref{eq:is}),
which
equals
\bqq
I_{S} = - {{64 \pi i} \over {15}} {{k^{4}} \over {\Omega^{4}}}
B(\phi_{1},\phi_{2},\phi_{1}^{'},\phi_{2}^{'}) \, . \label{eq:iss}
\eqq
Now for large $k$, the leading power of $k$ in $I_{V}$ is $k^2$. The dominant
contribution to the scattering amplitude will therefore come from $I_{S}$,
which behaves as $k^4$. However, on the side of large $k$ we are still
restricted by the condition (\ref{eq:klla}), and the validity of expressions
(\ref{eq:iv}) and (\ref{eq:iss}) is thus limited, say, to $2 < k < 10$.
For this interval of $k$, we have numerically checked the relative
contributions
of $I_{V}$ and $I_{S}$, and really found $I_{V}$ to be negligible. The final
result is therefore given by (\ref{eq:iss}).

\nonum
\section{Acknowledgements}

One of us (E.T.) wishes to acknowledge an early discussion with F. De Martini,
and informative discussions about lasers and synchrotron sources with S.
Modesti
and R. Rosei. We would like to acknowledge constructive remarks by an anonymous
referee. We also acknowledge support from INFM, and from the European
Research Office, U.S. Army.

\figure{The vacuum photon--photon fourth--order scattering diagrams.
\label{fig:1}}

\figure{Second order photon--photon scattering processes in a
nonrelativistic electron gas. \label{fig:3}}

\figure{A diagram with zero amplitude. \label{fig:4}}

\figure{Screening of the bare polarizability in the second order diagram.
\label{fig:5}}

\figure{Photon--photon planar scattering geometry (chosen for simplicity
of illustration). $\vec{k}_{1}, \vec{k}_{2}$ incoming photons,
$\vec{k}_{1}^{'}, \vec{k}_{2}^{'}$ outgoing photons. \label{fig:7}}

\figure{Differential cross section versus angle $\beta$ for the 2nd order
processes in the planar geometry of Fig. \ref{fig:7}, and for
$\theta = \pi/2$. The peaks are due to the plasmon resonance.
Their sharpness is a consequence of the random--phase approximation,
where the plasmon lifetime is infinite. \label{fig:6}}

\figure{A log--log plot of the calculated differential cross section versus
photon frequency for the 2nd order elastic photon--photon scattering processes
in an electron gas (away from plasmon resonances). Parameters chosen are
representative for potassium metal. \label{fig:10}}

\figure{"Triangle" diagrams corresponding to the 3rd order processes.
\label{fig:11}}

\figure{An overall graphical comparison of the photon--photon elastic
cross section in an electron gas and in vacuum. In an actual solid, or
liquid, both contributions will be present. \label{fig:12}}

\figure{Three diagrams corresponding to the 4th order processes in an
electron gas for the particular case of the scattering of two incident
photons with opposite wavevectors. \label{fig:8}}

\figure{The integration region, which is a thin outer shell of the
Fermi sphere. \label{fig:9}}

\begin{table}
\begin{tabular}{|p{0.2in}|p{0.2in}|p{0.2in}|p{0.2in}|p{0.2in}|p{0.3in}
                         |p{0.2in}|p{0.2in}|p{0.2in}|p{0.2in}|p{0.3in}
                         |p{0.2in}|p{0.2in}|p{0.2in}|p{0.2in}|}
\multicolumn{5}{c}{Diagram 1} & \multicolumn{6}{c}{Diagram 2} &
\multicolumn{4}{c}{Diagram 3} \\
                         \cline{1-5} \cline{7-10} \cline{12-15}
  & 1 & 2 & 3 & 4 & & 1 & 2 & 3 & 4 & & 1 & 2 & 3 & 4 \\
                         \cline{1-5} \cline{7-10} \cline{12-15}
a & $+$ & $-$ & $+$ & $-$ && $+$ & $-$ & $+$ & $+$ && $+$ & $+$ & $-$ & $+$ \\
                         \cline{1-5} \cline{7-10} \cline{12-15}
b & $-$ & $+$ & $-$ & $+$ && $+$ & $+$ & $+$ & $-$ && $-$ & $+$ & $+$ & $+$ \\
                         \cline{1-5} \cline{7-10} \cline{12-15}
c & $-$ & $+$ & $+$ & $+$ && $+$ & $-$ & $+$ & $-$ && $-$ & $+$ & $-$ & $+$ \\
                         \cline{1-5} \cline{7-10} \cline{12-15}
d & $+$ & $+$ & $-$ & $+$ && $-$ & $-$ & $-$ & $+$ && $+$ & $-$ & $-$ & $-$ \\
                         \cline{1-5} \cline{7-10} \cline{12-15}
e & $-$ & $-$ & $+$ & $-$ && $-$ & $+$ & $-$ & $-$ && $-$ & $-$ & $+$ & $-$ \\
                         \cline{1-5} \cline{7-10} \cline{12-15}
f & $+$ & $-$ & $-$ & $-$ && $-$ & $+$ & $-$ & $+$ && $+$ & $-$ & $+$ & $-$ \\
                         \cline{1-5} \cline{7-10} \cline{12-15}
g &  &  &  &  && $-$ & $+$ & $+$ & $+$ &&  &  &  &  \\
                         \cline{1-5} \cline{7-10} \cline{12-15}
h &  &  &  &  && $-$ & $-$ & $+$ & $+$ &&  &  &  &  \\
                         \cline{1-5} \cline{7-10} \cline{12-15}
i &  &  &  &  && $-$ & $+$ & $+$ & $-$ &&  &  &  &  \\
                         \cline{1-5} \cline{7-10} \cline{12-15}
j &  &  &  &  && $-$ & $-$ & $+$ & $-$ &&  &  &  &  \\
                         \cline{1-5} \cline{7-10} \cline{12-15}
k &  &  &  &  && $+$ & $+$ & $-$ & $+$ &&  &  &  &  \\
                         \cline{1-5} \cline{7-10} \cline{12-15}
l &  &  &  &  && $+$ & $-$ & $-$ & $+$ &&  &  &  &  \\
                         \cline{1-5} \cline{7-10} \cline{12-15}
m &  &  &  &  && $+$ & $+$ & $-$ & $-$ &&  &  &  &  \\
                         \cline{1-5} \cline{7-10} \cline{12-15}
n &  &  &  &  && $+$ & $-$ & $-$ & $-$ &&  &  &  &  \\
                         \cline{1-5} \cline{7-10} \cline{12-15}
\end{tabular}
\caption[...]{\it Terms giving a non-zero contribution to the diagrams 1-3.
		    $+$ denotes a hole, $-$ an electron.}
\label{t1}
\end{table}

\end{document}